\documentclass[twocolumn,preprintnumbers,amsmath,amssymb]{revtex4}

\usepackage{graphicx}
\usepackage{dcolumn}
\usepackage{bm}
\usepackage{epstopdf}

\begin{document}

\title{Mitigating the effect of non-uniform loss on time reversal mirrors}

\author{Biniyam Tesfaye Taddese\textsuperscript {1, 2}}
\author{Thomas M. Antonsen\textsuperscript {2, 3}}
\author{Edward Ott\textsuperscript {2, 3}}
\author{Steven M. Anlage\textsuperscript {1, 2}}
\affiliation{\textsuperscript {1}Center for Nanophysics and Advanced Materials, Department of Physics, University of Maryland, College Park, Maryland 20742-4111, USA.
\textsuperscript {2}Department of Electrical and Computer Engineering, University of Maryland, College Park, Maryland 20742-3285, USA.
\textsuperscript {3}Department of Physics, University of Maryland, College Park, Maryland 20742-4111, USA.
}

\date{\today}

\begin{abstract}
Time reversal mirrors work perfectly only for lossless wave propagation. Here, the performance of time-reversal mirrors is quantitatively defined, and the adverse effect of dissipation on their performance is investigated. An application of the technique of exponential amplification is proposed to overcome the effect of dissipation in the case of uniform loss distributions, and, to some extent, in the case of non-uniform loss distributions. A numerical model of a star graph was employed to test the applicability of this technique on realizations with various random spatial distributions of loss. A subset of the numerical results are also verified by experimental results from an electromagnetic time-reversal mirror.  The exponential amplification technique should improve the performance of emerging technologies based on time-reversed wave propagation such as directed communication and wireless power transfer.
\end{abstract}
\maketitle

\clearpage

\section{Introduction \label{sec:5-Introduction}}
	Time reversal (TR) mirrors can, under ideal circumstances, precisely reconstruct a wave disturbance which happened at an earlier time, at any given later time. They have found numerous applications since the earliest experimental demonstrations of TR using acoustic waves \cite{Fink1996}. More recently, TR mirrors have been realized using electromagnetic waves \cite{Lerosey2004, Anlage2007} expanding their range of applicability. They have applications in communications \cite{Xiao2007, Lerosey2005, Frazier2013a, Frazier2013b}, imaging \cite{Liu2005, Chen2008}, source localization \cite{Larmat2009}, non-destructive evaluation \cite{Ulrich2008, Ulrich2009}, selective beamforming \cite{Sun2013}, and sensing \cite{Taddese2009, Taddese2010, Anlage2007}. The principle of TR mirrors is closely linked with the concepts of scattering fidelity \cite{Gorin2006}, and the Loschmidt echo \cite{Taddese2009}.

	TR mirrors are used to focus waves both in space and time. An ideal TR mirror consists of a wave source located inside a lossless medium that is completely enclosed by a surface of transceivers. A TR mirror operates in two steps. In step one, the transceivers record and absorb the signal broadcast by the source. In step two, the transceivers rebroadcast time reversed versions of the recorded waves. The waves broadcast in step two eventually focus on the location of the source and reconstruct a time reversed version of the original signal, which was broadcast in step one. This is possible because of the TR invariance property of the lossless wave equation. This ideal TR mirror uses the so called closed TR cavity \cite{Cassereau1992}. However, it is not generally practical, for example, to build a closed TR cavity, whose interior is completely covered with transceivers. Practical TR mirrors have several limitations, which result in an imperfect reconstruction of the original signal. These limitations include i) limited spatial coverage by the transceivers, and ii) dissipation during the wave propagation (which breaks TR invariance of the wave equation) \cite{Draeger1997, Taddese2009}.

	The first limitation of TR mirrors can be overcome by the use of a reflecting wave chaotic cavity with partial spatial coverage by transceivers that have a long recording time \cite{Draeger1997, Taddese2009}. However, the limitation due to dissipation persists, and leads to increasing loss of information as the recording time increases.

	The technique of iterative TR was proposed to improve the performance of TR mirrors which suffer from the limitations of dissipation and incomplete spatial coverage of transceivers \cite{Montaldo2004}. However, the iterative technique may not converge fast enough, despite a recent result showing the tunability of its convergence \cite{Taddese2011}. A faster alternative to the iterative technique is the inverse filter technique \cite{Tanter2000, Montaldo2004}. But, the inverse filter technique is computationally costly and potentially unstable \cite{Montaldo2004}.

On the other hand, the technique of exponential amplification was used to overcome the adverse effect of dissipation on TR based sensing techniques \cite{Taddese2009, Taddese2010}. The spatial range of these techniques is limited due to dissipation. The exponential amplification technique was successfully employed to extend the spatial range of the TR based sensing techniques.  Exponential amplification has also been employed to remove the effects of dissipation on the scattered field intensity in disordered wave propagation systems \cite{Chabanov2000}.

In this paper, another application of the exponential amplification technique is proposed. Unlike the work in Ref.\cite{Taddese2010}, the application involves improving the performance of TR mirrors which suffer from dissipation. The performance of TR mirrors is defined as the normalized correlation between the original pulse and the time reversed version of the reconstructed pulse (after some post-processing). This paper quantitatively demonstrates that the exponential amplification technique mitigates the adverse effects of dissipation on the performance of TR mirrors, for different spatial loss distributions. The technique of exponential amplification can compete with the iterative TR technique in improving this performance of TR mirrors, to some extent. The advantage of the exponential amplification technique is that it is faster to implement, as it does not rely on iterative steps. It is also cheaper computationally. The application of the exponential amplification proposed in this paper can enable better TR mirror based communication and wireless power transfer systems in a lossy environment, as discussed in Sec.~\ref{sec:5-Discussion}.

In Sec.~\ref{sec:5-Theory}, the theory supporting the exponential amplification technique, first discussed in Ref.\cite{Taddese2010}, is summarized. The performance of TR mirrors, which was not discussed in Ref.\cite{Taddese2010}, is also defined in Sec.~\ref{sec:5-Theory}. In Sec.~\ref{sec:5-Experiment}, the exponential amplification technique is used to improve the newly defined performance of an electromagnetic TR mirror. In particular, it is experimentally shown that the technique can improve the performance of a TR mirror for the case of spatially uniform distributed loss. Sec.~\ref{sec:5-Simulation} investigates the limits of applicability of the exponential amplification in systems with inhomogeneous loss distributions. Sections ~\ref{sec:5-Discussion} and ~\ref{sec:5-Conclusion} present further discussion of application of these ideas, and conclusions.

\section{Theory \label{sec:5-Theory}}
The theory supporting the exponential amplification technique was originally presented in Ref.\cite{Taddese2010}. In this section, the theory is summarized and a quantitative measure of performance of TR mirrors is defined. To summarize the theory, a cavity with transfer function $s(\omega)$ is considered. When a brief pulse, $a(t)$, is broadcast into the cavity, the response is $b(t)$. At a later time, a time reversed version of $b(t)$ denoted $b(-t)$ is broadcast back into the cavity producing reconstructed pulse denoted $c(-t)$ at the original broadcasting antenna or transducer.

Ref.\cite{Taddese2010} provides a derivation showing that, if the cavity is lossless, $a(t)$ is the same as a time reversed version of $c(-t)$ denoted here as $c(t)$. Thus, the TR mirror works perfectly in a lossless cavity.

Even if the cavity is lossy, it is shown that the TR mirror can still recover the original pulse in the special case of uniform loss. Suppose that the uniform loss is represented by a $1/e$ amplitude decay time $\tau$. In this case, the signal $b(t)$ is exponentially amplified by $e^{2t/\tau}$, time reversed, and broadcast back into the cavity. $\tilde{b}(-t)$ denotes the signal obtained after exponentially amplifying $b(t)$, and time reversing it. Similarly, $\tilde{c}(-t)$ denotes the reconstructed pulse obtained when $\tilde{b}(-t)$ is broadcast back into the cavity. Ref.\cite{Taddese2010} analytically predicts that $\tilde{c}(t)e^{-2t/\tau}$ is equal to $a(t)$ for a uniform loss that is characterized by $\tau$.

Here, we define a new signal $\hat{c}(t)$ which is equal to $\tilde{c}(t)e^{-2t/\tau}$. $\hat{c}(t)$ is called the corrected reconstructed pulse. In this paper, the performance ($\eta$) of the TR mirror is defined based on the normalized correlation between the corrected reconstructed pulse ($\hat{c}(t)$), and the original pulse ($a(t)$).
\begin{equation}
\eta \equiv \frac{\sum_{t=0}^{t=T}a(t)\hat{c}(t)}{\sqrt{\sum_{t=0}^{t=T}a(t)^{2} \sum_{t=0}^{t=T}\hat{c}(t)^{2} }}
\label{eqn:TRperf}
\end{equation}
Each of the signals $a(t)$ and $\hat{c}(t)$ have a duration of $T$. In principle, $\eta$ can range from $-1$ to $1$. However, since the two signals are aligned to maximize $\eta$, $\eta$ ranges from $0$ (poor performance) to $1$ (perfect reconstruction). Based on the result from Ref.\cite{Taddese2010}, $a(t)=\hat{c}(t)$ if the system is uniformly lossy with a $1/e$ voltage decay time, $\tau$; therefore, the performance of the TR mirror is expected to be perfect (i.e. $\eta=1$). However, the analysis does not model the effect of noise and limited dynamic range on the exponential amplification technique.

If the exponential amplification is not applied to $b(t)$, the performance, $\eta$, can still be calculated using the signals $a(t)$ and $c(t)$, instead of using $a(t)$ and $\hat{c}(t)$. But, if the system is lossy, and exponential amplification is not applied, then it is not expected that $\eta=1$.

In Sec.~\ref{sec:5-Experiment}, this theory, which assumes uniform loss distribution, is successfully tested in an experimental system that approximates the case of uniform loss distribution. However, it is also expected that the technique might be useful in the case of a moderately non-uniform loss distribution. The applicability of the exponential amplification technique to mitigate the effect of non-uniform loss is studied in Sec.~\ref{sec:5-Simulation}.

\section{Overcoming the Effect of Spatially Uniform Loss: Experimental Test \label{sec:5-Experiment}}
\begin{figure}
\begin{center}
\includegraphics[width=3.5in]{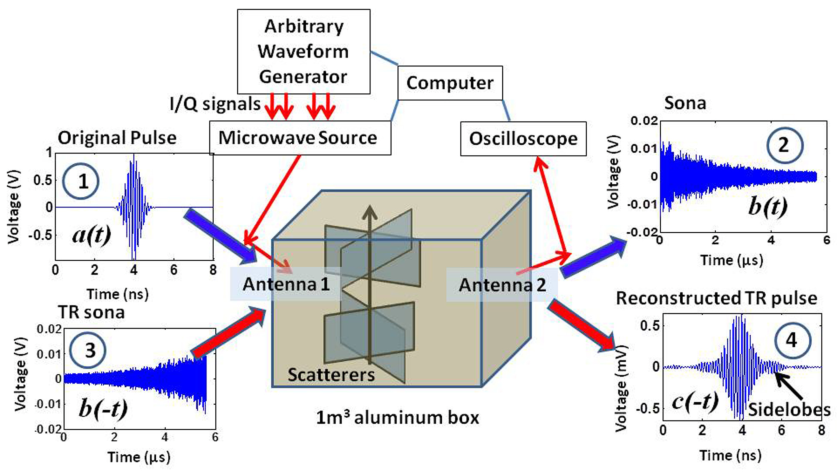}
\end{center}
\begin{quote}
\caption{\label{fig:eaFig1}
Schematic of the electromagnetic time reversal (TR) mirror experiment, without exponential amplification.
During the first step of the TR mirror, the original pulse is broadcast through antenna 1
(as shown in part 1), and the resulting sona is collected at antenna 2 (as shown in part 2).
During the next step of the TR mirror, the time reversed sona is injected into the system at antenna 1 (as shown in part 3) to retrieve the reconstructed time reversed pulse at antenna 2, making use of spatial reciprocity (as shown in part 4).
Experimental data are shown for each step.}
\end{quote}
\end{figure}

The electromagnetic TR mirror operates as shown in Fig.~\ref{fig:eaFig1}. The TR mirror involves a one cubic meter aluminum box that has irregular scatterers inside it, and two ports in its walls. The aluminum walls of the cavity are constructed of a single material, and can be assumed to have a uniform spatial distribution of loss. The original pulse was broadcast into the cavity using port 1, and the response signal, which we call "sona", was collected using port 2. The original pulse had a carrier frequency of $7 GHz$ and a Gaussian envelope with a standard deviation of $1 ns$. The sona was recorded for about $6.5 \mu s$ with a signal to noise ratio (SNR) greater than $1$. During the second step of the TR mirror operation, the sona signal was time reversed and broadcast back into the cavity using port 1; this made use of the spatial reciprocity property of the wave equation. Finally, a reconstructed pulse, which approximates a time reversed version of the original pulse was collected at port 2. The reconstructed pulse (and its temporal sidelobes) was collected over a $10 \mu s$ duration. The first $6.5 \mu s$ included a direct recording of the time reversed sona being injected into the cavity from port 1. The last $3.5 \mu s$ was a recording of the post-reconstructed pulse emerging from port 2, reverberating throughout the cavity.

Based on Eq.~\ref{eqn:TRperf}, $\eta$ can be calculated as the normalized correlation between the following two aligned signals: i) the time reversed version of the reconstructed pulse with its temporal sidelobes (which was a $10\mu s$ long recording), and ii) the original pulse broadcast (by considering the $1ns$ long Gaussian envelope to be located at $t=3.5\mu s$ of a $10\mu s$ long signal of mostly zero voltage values). Note that these two signals are equivalent to the aligned $a(t)$ and $c(t)$ signals introduced in Sec.~\ref{sec:5-Theory}. If the TR mirror were perfect, $\eta$ would be $1$. However, as can be seen in part 4 of Fig.~\ref{fig:eaFig1}, the reconstructed pulse has temporal sidelobes which result in $\eta<1$. The sidelobes arise due to the limitations of practical TR mirrors discussed in Sec.~\ref{sec:5-Introduction}, which include limited spatial coverage of transceivers and dissipation.

Generally, the sona signal shows an exponential decay which is caused by dissipation and coupling losses. The experimental system discussed here is strongly under-coupled \cite{Hemmady2006}, which means that the effect of dissipation dominates the effect of coupling. Therefore, the $1/e$ voltage decay constant ($\tau$) which is related to the unloaded quality factor can be readily determined by a linear fit to the $log$ of the sona energy as a function of time. The exponential amplification can then be applied to the sona signal before it is time reversed and broadcast into the cavity. The exponential amplification is carried out by multiplying the sona signal by the time dependent amplifying function
\begin{equation}
A(t,F) = exp(\frac{Ft}{\tau})
\end{equation}
where $t$ is time in seconds, and $F$ is an adjustable factor. Assuming a precise determination of $\tau$, using $F=1$ compensates only for the effect of dissipation on the sona during step 1 of the TR mirror operation; whereas, using $F=2$ compensates for the effect of dissipation on the sona during both step 1 and step 2 of the TR mirror operation \cite{Taddese2010}. The theoretical discussion in Sec.~\ref{sec:5-Theory} predicts that $F=2$ to maximize $\eta$.  The exponential amplification is applied to the part of the sona whose signal to noise ratio is, at the very least, greater than $1$ (i.e. the $6.5\mu s$ long sona in this experiment). In addition, the function $A(t,F)$ is typically terminated by a smooth ramp-down function whose time span is at least as wide as the time duration of the original pulse, which is $1ns$; this prevents additional frequency components from entering into the sona.

\begin{figure}
\begin{center}
\includegraphics[width=3.5in]{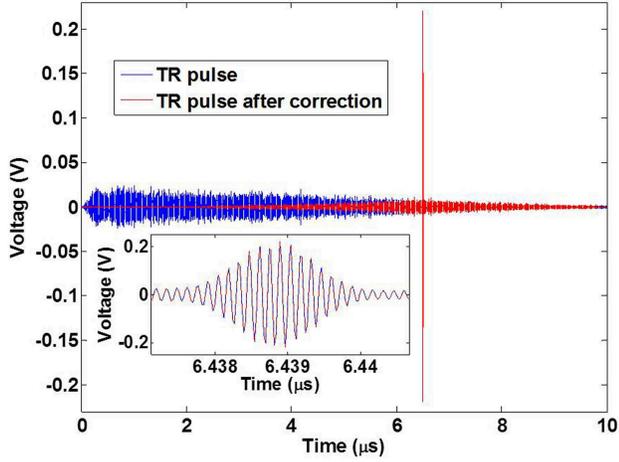}
\end{center}
\begin{quote}
\caption{\label{fig:eaFig2}
The reconstructed TR pulse obtained from a sona that was exponentially amplified with $F=2$ (blue) had significant sidelobes. The corrected reconstructed pulse (red) was obtained by multiplying the time reversed version of the reconstructed TR pulse by $A(t-3.5\mu s, F=-2)$. The corrected reconstructed pulse is displayed here after time reversing it. $A(t,F)=1$ if $t<0$. The inset shows a close up of the reconstructed pulses, which are the same before and after the correction.}
\end{quote}
\end{figure}

Suppose that the sona is exponentially amplified using $A(t,F)$, before it is time reversed and broadcast into the electromagnetic cavity. The reconstructed pulse obtained with $F>0$ will have more sidelobes than the reconstructed pulse obtained with $F=0$. However, these sidelobes can be numerically suppressed by manipulating the reconstructed waveform as follows. Fig.~\ref{fig:eaFig2} shows the reconstructed waveform obtained using sona that is amplified with $A(t,F=2)$ (in blue); the reconstructed waveform has unwanted temporal sidelobes which are a result of the exponential amplification applied to the sona. These sidelobes were corrected as follows. The $10\mu s$ long reconstructed waveform was time reversed so that the $1ns$ long Gaussian pulse is located at about $t=3.5\mu s$; this waveform was then multiplied by $A(t-3.5\mu s, F=-2)$ to get the corrected reconstructed waveform (i.e. $\hat{c}(t)$) introduced in Sec.~\ref{sec:5-Theory}. $A(t,F)=1$ for $t<0$ by convention. In Fig.~\ref{fig:eaFig2}, the corrected reconstructed waveform is time reversed and plotted (in red) to highlight its relationship with the reconstructed waveform. The corrected reconstructed waveform is expected to have a better normalized correlation with the original pulse because of the exponential amplification with $F=2$, as predicted in Sec.~\ref{sec:5-Theory}.

\begin{figure}
\begin{center}
\includegraphics[width=3.5in]{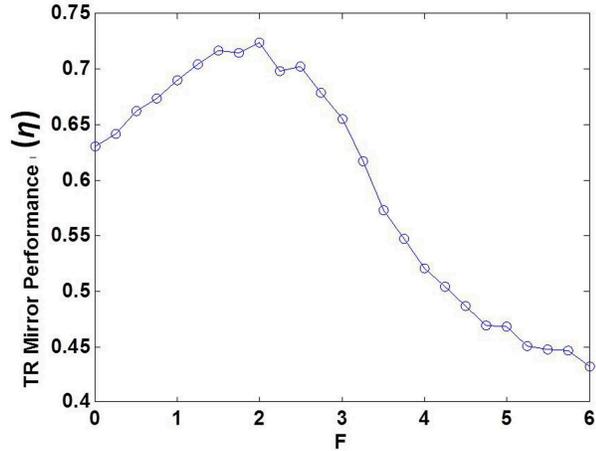}
\end{center}
\begin{quote}
\caption{\label{fig:eaFig3}
The TR mirror performance ($\eta$) of the experimental electromagnetic TR mirror as a function of the $F$ parameter used to exponentially amplify the sona signal. The optimum value of the parameter is $F=2$.}
\end{quote}
\end{figure}

Several values of $F$ were used to carry out the operation of the TR mirror assisted by exponential amplification. For each $F$ value, the performance of the electromagnetic TR mirror, $\eta$, is shown in Fig.~\ref{fig:eaFig3}. The performance of the TR mirror is enhanced the most by using exponential amplification with $F\approx2$. In particular, $\eta$ increases from $\approx63\%$ at $F=0$ to $\approx73\%$ at $F=2$. This result agrees with the theoretical prediction in Sec.~\ref{sec:5-Theory}. The fact that the $\eta<1$ even for $F\approx2$ can be explained by the fact that the TR mirror has only one recording channel, whereas the theory assumes that all the waves are captured. Thus, the exponential amplification technique only overcomes the limitation of the TR mirror associated with dissipation, and not the problem of insufficient spatial coverage of recorders. Noise, which is not considered in the theory, also affects the performance of TR mirrors as seen in Ref.\cite{Taddese2011}.

As shown in Fig.~\ref{fig:eaFig3}, $\eta\approx63\%$ if the exponential amplification technique is not applied. The exponential amplification with the optimum $F=2$ parameter achieves $\eta\approx73\%$ for the experimental set up described. This compares with $\eta\approx78\%$ which is achieved by the tunable iterative technique on the same experimental set up \cite{Taddese2011}. Therefore, the iterative technique performs better than the exponential amplification technique when both of them use their respective optimum parameters. Nonetheless, the exponential amplification has an advantage because of its speed, and computational simplicity.

The result in this section proves that the exponential amplification improves the performance, $\eta$, of TR mirrors with uniform spatial loss.  The results are also in good qualitative agreement with expectations for exponential amplification.

\section{Overcoming the Effect of Spatially Non-Uniform Loss: Numerical Test \label{sec:5-Simulation}}

The exponential amplification technique is derived assuming uniform loss distribution. The case of non-uniform loss distribution is better handled with techniques such as the inverse filter, and the iterative TR \cite{Tanter2000, Montaldo2004}, which do not assume uniformity of loss. However, it is expected that the exponential amplification can be successfully applied in cavities with moderately inhomogeneous loss distributions.

We have repeated the experiments discussed in Sec.\ref{sec:5-Experiment} inside a similar cavity with a non-uniform spatial loss distribution, and we have shown that $\eta$ increases from $\approx47\%$ at $F=0$ to $\approx52\%$ at $F=2$.  This modest improvement is due to the use of very lossy microwave absorber material to vary the loss, resulting in very inhomogeneous loss distribution.  In this section, a systematic and better controlled numerical study of the effect of non-uniform spatial loss distribution on the applicability of the exponential amplification technique is presented.

The applicability of the exponential amplification technique was studied as the inhomogeneity of the spatial distribution of loss was increased. The simulations in this section were designed to supplement the experimental results in Sec. \ref{sec:5-Experiment}.

\subsection{Simulation Setup \label{sec:5-SimuSetUp}}
\begin{figure}
\begin{center}
\includegraphics[width=3in]{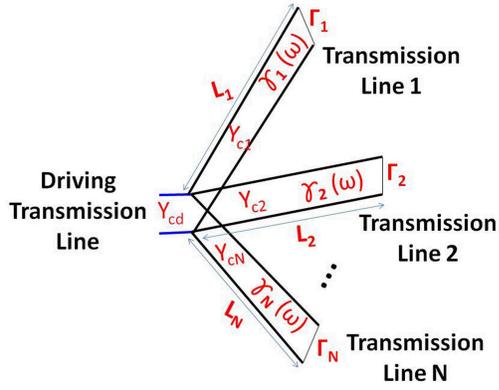}
\end{center}
\begin{quote}
\caption{Schematic of the one-port star graph model shows $N$ transmission lines and a driving transmission line, which are all connected in parallel. Each of the $N$ lines (labeled by $i$) can have unique length ($L_{i}$), characteristic admittance ($Y_{ci}$), frequency dependent propagation constant ($\gamma_{i}(\omega)$), and termination reflection coefficient ($\Gamma_{i}$). The driving line has a characteristic admittance of $Y_{cd}$ and zero length.  \label{fig:eaFig4}}
\end{quote}
\end{figure}

The numerical model is an implementation of a star graph \cite{Kottos2003, Taddese2011, Frazier2013a, Frazier2013b}. We use the star graph because it is a type of quantum graph that has generic properties of wave chaotic systems \cite{Kottos2003}, but it is relatively simple to understand and simulate. The star graph is a quasi-1D chaotic system. It is modeled using a driving transmission line that feeds a number of other transmission lines, which are all connected to each other in parallel at a single node (see Fig.~\ref{fig:eaFig4}). Each transmission line is terminated with a reflection load. The graph is a one port system. Thus, the original pulse is injected through the driving transmission line, and the sona is also collected from the same line. The original pulse has the same characteristics as the one used in the experiment described in Sec.~\ref{sec:5-Experiment}.

There are $500$ transmission lines in the star graph, in addition to the driving line. Their lengths are given by $L_{i}=\sqrt{i}$ $m$ for integer $i$ ranging from $1$ to $500$. The driving line has a length of $0$ $m$. The characteristic admittances of the $500$ lines are $Y_{ci}=1S$. The driving line has a characteristic admittance of $Y_{cd}=\sum_{i=1}^{i=500}Y_{ci}=500S$; this choice eliminates prompt reflection of signals injected through the driving line. The terminal reflection coefficients, $\Gamma_{i}$, of the $500$ lines are all set to $1$. The dissipation is introduced through the frequency dependent propagation constant of the lines, $\gamma_{i}(\omega)=j\frac{\omega}{c}+\alpha_{i}$, where $j$ is the square root of $-1$, $c$ is the speed of light and $\alpha_{i}$ is the loss constant of line $i$.

\begin{figure}
\begin{center}
\includegraphics[width=3.5in]{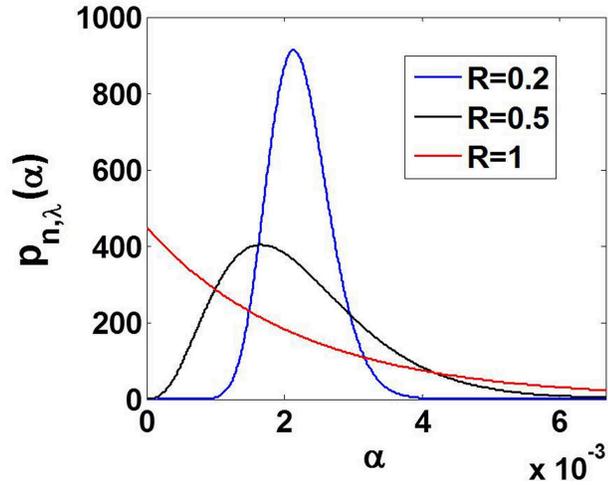}
\end{center}
\begin{quote}
\caption{\label{fig:eaFig5}
The probability density function (pdf) of propagation loss (Eq.~\ref{eqn:5-LossPDF}) is plotted for different $R=\sigma_{\alpha}/\mu_{\alpha}$ values where $\mu_{\alpha}\approx0.002$. $R$ is a measure of loss inhomogeneity with larger values representing systems with wider loss variation. The pdf is plotted for $R=0.2$ (blue), $R=0.5$ (black), and $R=1$ (red).}
\end{quote}
\end{figure}

$\alpha_{i}$ specifies that the voltage waves decay on line $i$ as $e^{-\alpha_{i}z}$, where $z$ is distance measured along line $i$ in meters. The spatial inhomogeneity of loss on the star graph is modeled as follows. The $\alpha_{i}$ of each of the $500$ lines is randomly chosen from the probability density function (pdf),
\begin{equation}
p_{n,\lambda}(\alpha) = \frac{\alpha^{n}}{\lambda^{n+1}n!}exp(-\alpha/\lambda) .
\label{eqn:5-LossPDF}
\end{equation}
The pdf has two parameters, $n$ and $\lambda$, which define the mean ($\mu_{\alpha}=\lambda+\lambda n$) and standard deviation ($\sigma_{\alpha}=\sqrt{\lambda(\lambda+\lambda n)}$) of $\alpha$ values. This particular pdf is chosen for the following two reasons. First, the coefficient of variation ($R=\sigma_{\alpha}/\mu_{\alpha}$) can be easily varied while keeping $\mu_{\alpha}$ constant. Keeping the average spatial loss constant simplifies the problem, and helps us focus on the effect of increasing spatial loss inhomogeneity. The main motivation for using the exponential amplification in the case of non-uniform loss is this: if the $1/e$ decay constants (i.e. $\tau$) of the different modes are not extremely different, we can use a single average $\tau$ value. Thus, it is not interesting to vary the average loss (effectively $\tau$) here. In this set up $\mu_{\alpha}$ is chosen as $(c\tau)^{-1}$, where $\tau=1.5\mu s$. The second reason to use this particular pdf is that its support is the set of positive numbers, which should be the case as $\alpha$ should always be positive on the passive transmission lines. Fig.~\ref{fig:eaFig5} shows plots of the pdf for different values of $R$, where $\mu_{\alpha}\approx0.002$ is constant.

The numerical values of the parameters of the star graph were chosen to achieve the following objective. The experimental result in Sec.~\ref{sec:5-Experiment} was based on an under-coupled cavity. An under-coupled cavity is characterized by the domination of energy loss due to dissipation over energy loss due to coupling \cite{Hemmady2006}. If the cavity is not under-coupled, the advantage of the exponential amplification technique cannot be easily seen because there are no strong internal dissipation effects to be compensated in an over-coupled cavity. To accomplish under-coupling, $500$ lines were used in the star graph to increase the back-reflection coefficient of the waves that are trying to leave the star graph. This forced the waves to reverberate inside the star graph longer, which decreased the coupling loss by two orders of magnitude compared to the dissipation loss.

\subsection{Simulation Results \label{sec:5-SimuSetUp}}

A TR mirror was implemented on the star graph by broadcasting the original pulse, collecting a sona, exponentially amplifying the sona with $A(t,F)$, time reversing the amplified sona, and broadcasting it back into the star graph. The reconstructed pulse was collected, and it was time reversed and multiplied by $A(t-t_{p},-F)$. Here, $t_{p}$ is the time when the $1ns$ Gaussian pulse was expected within the time reversed version of the reconstructed waveform. The signals were collected over a span of $10 \mu s$ without noise.

\begin{figure}
\begin{center}
\includegraphics[width=3.5in]{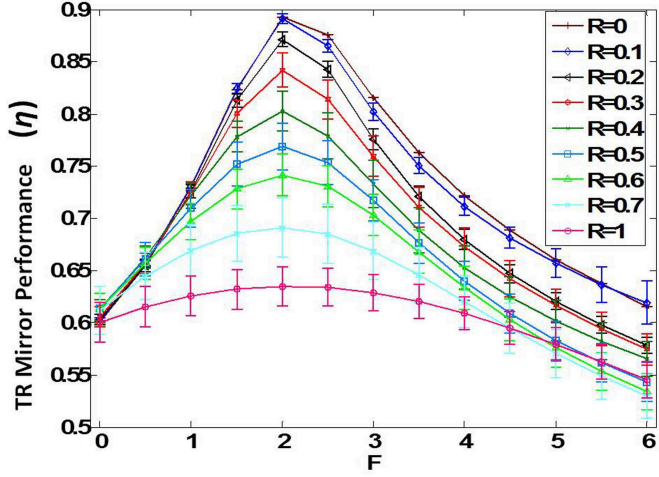}
\end{center}
\begin{quote}
\caption{\label{fig:eaFig6}
The performance ($\eta$) of the TR mirror in a star graph model as a function of $F$ parameter for various degrees of loss inhomogeneity characterized by $R$. Loss inhomogeneity increases with $R$. The optimum $F$ value is always around $F=2$ for the $\tau$ value that is well determined from the sona.}
\end{quote}
\end{figure}

The performance ($\eta$) of the TR mirror is plotted versus $F$ as shown in Fig.~\ref{fig:eaFig6}. This entire process was repeated for star graphs with different degree of loss inhomogeneity characterized by $R$. $R=0$ represents uniform spatial loss distribution. As $R$ increases, the loss inhomogeneity increases. For each $R$ value, $25$ realizations of the star graph were generated; hence, there are error bars included on the $\eta$ vs $F$ plots for each $R$ value, as shown in Fig.~\ref{fig:eaFig6}.

\begin{figure}
\begin{center}
\includegraphics[width=3.5in]{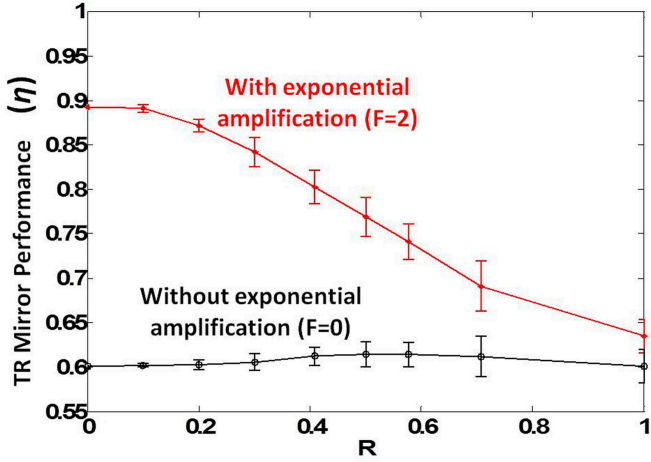}
\end{center}
\begin{quote}
\caption{\label{fig:eaFig7}
The performance ($\eta$) of the TR mirror in a star graph model is plotted as a function of the loss inhomogeneity, $R$, for the cases of: i) no exponential amplification (shown in black), and ii) optimal exponential amplification, which is always $F=2$ (shown in red). It is clear that exponential amplification is advantageous for uniform and moderately non-uniform spatial loss distributions.}
\end{quote}
\end{figure}

For each $R$ value shown in Fig.~\ref{fig:eaFig6}, the maximum $\eta$ was taken over all $F$ values attempted; it turns out that the maximum $\eta$ is always achieved for $F=2$. The maximum $\eta$ achieved is plotted as a function of the loss inhomogeneity ($R$) in Fig.~\ref{fig:eaFig7} (shown in red). The $\eta$ achieved without using exponential amplification (i.e. $F=0$) is also plotted as a function of $R$ in Fig.~\ref{fig:eaFig7} (shown in black). Fig.~\ref{fig:eaFig7} demonstrates that the exponential amplification technique improves $\eta$ significantly, for small values of $R$, which are close to uniform loss distributions. The exponential amplification is still applicable in cavities with more inhomogeneous loss distributions. However, its effectiveness declines with increasing loss inhomogeneity as expected. The results are consistent with the experimental observations on a system with inhomogeneous loss, mentioned at the beginning of this section.

From Fig.~\ref{fig:eaFig7}, we can see that performance $\eta$ is not affected significantly by loss inhomogeneity $R$ if exponential amplification is not applied. Here, it is worth repeating the fact that the average spatial loss in the star graph is constant for all the $R$ values considered. On the other hand, the fact that $\eta$ is not restored to $1$ even for $R=0$ and $F=2$ can be explained by the fact that the star graph is designed to be under-coupled, to match the nature of the experimental cavity in Sec.~\ref{sec:5-Experiment}. The waves in the star graph will continue to reverberate for longer than the $10\mu s$ sona recording time. Thus, the problem is not just dissipation which can be compensated by the exponential amplification technique, but also the lack of sufficient temporal coverage, which has previously been shown to affect the quality of TR focusing \cite{Draeger1999}.

\section{Discussion \label{sec:5-Discussion}}
The tunable iterative time-reversal technique overcomes the adverse effect of dissipation on the real-time spatiotemporal wave focusing of time reversal mirrors \cite{Taddese2011}. The focusing was quantified using the ratio of pulse to temporal sidelobe energy. In this paper, we are less interested in achieving a real-time spatiotemporal wave focusing using time reversal mirrors. Instead we are mainly interested in improving the quality of the reconstructed waveforms to more nearly reproduce the original waveform.  We have demonstrated that exponential amplification can significantly improve the reconstruction quality even in the presence of moderately inhomogeneous loss.  The ability to improve reconstructed signal fidelity can be useful. For instance, if the shape of the reconstructed pulse is encoding information in a communication application, \cite{Frazier2013a, Frazier2013b} it is important to be able to improve the fidelity of the reconstructed pulse (which is measured by $\eta$) by suppressing its temporal sidelobes.  In addition, in wireless power transfer technology utilizing time-reversed waveforms, it will be necessary to optimize energy delivery by tailoring the signal to match the energy rectifying element.\cite{Frazier2013a}

\section{Conclusion \label{sec:5-Conclusion}}
Exponential amplification improves the performance of time reversal mirrors best if the loss is uniformly distributed in space. However, even under conditions in which the loss is not uniformly distributed, the exponential amplification mitigates the adverse effect of non-uniform loss on the performance of time reversal mirrors. The exponential amplification works best when the theoretically predicted optimum parameter of $F=2$ is used. This technique may be advantageous to use in time reversal mirror based communication applications because even if the reconstructed pulse will have significant temporal sidelobes during the recording, the processed reconstructed pulse will closely replicate the shape of the original pulse.

This work is supported by ONR grant N00014130474, ONR MURI grant N000140710734, AFOSR grant FA95500710049, ONR AppEl Center, Task A2, grant N000140911190, and the Maryland Center for Nanophysics and Advanced Materials.

\clearpage

\end{document}